\documentclass[11pt,a4paper]{article}
\pdfoutput=1
\usepackage{jcappub}

\usepackage{courier}
\usepackage{appendix}
\usepackage{array}
\newcolumntype{x}[1]{>{\centering\arraybackslash}p{#1}}

\usepackage{epsfig}
\usepackage{graphicx}
\usepackage{dcolumn}
\usepackage{amsmath}
\usepackage{enumerate}
\def\lsim{\mathrel{\hbox{\rlap{\hbox{\lower4pt\hbox{$\sim$}}}\hbox{$<$}}}}
\DeclareMathOperator{\erf}{erf}

\DeclareSymbolFont{matha}{OML}{txmi}{m}{it}
\DeclareMathSymbol{\varv}{\mathord}{matha}{118}

\makeatletter

\newcommand{\Rmnum}[1]{\expandafter\@slowromancap\romannumeral #1@}
\makeatother

\title{Spin-dependent dark matter interactions at loop-level in Ar and Xe}

\author[a,b]{Nassim Bozorgnia,}
\author[c]{Muping Chen,}
\author[d]{Graciela B. Gelmini,}
\author[d]{Alvine C. Kamaha,}
\author[d]{and Yongheng Xu}

\affiliation[a]{Department of Physics, University of Alberta,\\ 
Edmonton, Alberta T6G 2E1, Canada} 
\affiliation[b]
{Theoretical Physics Institute, University of Alberta,\\
Edmonton, Alberta T6G 2E1, Canada}
\affiliation[c]
{WPI-QUP, KEK, Oho 1-1, Tsukuba, Ibaraki 305-0801, Japan}
\affiliation[d]{Department of Physics and Astronomy, UCLA, \\
475 Portola Plaza, Los Angeles, CA 90095, USA}

\emailAdd{nbozorgnia@ualberta.ca}
\emailAdd{mpchen@post.kek.jp}
\emailAdd{gelmini@physics.ucla.edu}
\emailAdd{akamaha@physics.ucla.edu}
\emailAdd{xuyongheng@physics.ucla.edu}

\abstract{Xenon and argon are the two noble gases used in tonne scale dark matter direct detection experiments. We compare the detection capability of both target elements for interactions due to a pseudoscalar mediator including loop-level contributions to the cross section. At tree-level this type of interaction depends on the nuclear spin and would thus not be detectable in argon-based detectors, since Ar has spin zero. However, at the loop-level the same interaction yields spin-independent contributions that would be detectable in an argon target and are not negligible with respect to the tree-level interactions in xenon, because these are momentum suppressed. In fact, the loop-level contributions are also important for xenon-based experiments at low recoil energies, which could change their discovery reach for this interaction.}

\begin{document}
\maketitle

\section{Introduction}

Direct dark matter (DM) detection attempts to detect the energy deposited by the collision of a DM particle from the dark halo of our galaxy within a detector. Xenon (see e.g.~\cite{XENON:2019rxp, XENON:2020kmp, LZ:2018qzl, LZ:2019sgr, PandaX:2018wtu, PandaX-4T:2021bab, Baudis:2024jnk, aalbers2024xlzd})
and argon (see e.g.~\cite{DarkSide-20k:2017zyg, ArDM:2017ndf, DarkSide-50:2023fcw, Manthos:2023swh, Aalbers:2022dzr,Bonivento:2024qpn})
are the two liquefied noble gases used in tonne scale direct detection experiments. The proposed  40 tonne (50 tonne total) DARWIN~\cite{DARWIN:2016hyl}, 60 tonne (75 tonne total) XLZD~\cite{Baudis:2024jnk, aalbers2024xlzd} and 43 tonne (47  tonne total) PandaX-xT~\cite{PandaX-4T:2021bab} DM direct detection experiments with liquid xenon,  and  300 tonne Global Argon Dark Matter Collaboration (GADMC) consortium ARGO 
detector~\cite{Galbiati-2018, DarkSide20k:2020ymr} with liquid argon, will reach the sensitivity at which the interactions from coherent elastic neutrino-nucleus scattering -- the ``neutrino fog" --  will become the most important background.  

The best search strategy in direct DM detection is to cast as wide a net as possible in terms of DM-nucleus interactions (see e.g.~ref.~\cite{Gelmini:2018ogy}). Because the argon nucleus does not have spin, argon-based direct detection experiments are insensitive to all interactions coupling the DM particle to nuclear spin, such as those with axial vector or pseudoscalar mediators at tree-level. Xenon  is instead sensitive to these interactions. It is thus important to understand how much of a disadvantage argon-based detectors have with respect to those that are xenon-based to detect this type of interactions. The same interactions that lead to a nuclear spin-dependent cross section at tree-level,  at the loop-level always contain terms that are independent of the nuclear spin, and could thus be detected in argon. 

Loop-level corrections must include all the particles than can be exchanged within the loop given a particular particle model, and not only the main mediator that is usually sufficient to compute tree-level interactions. These corrections can be computed in a straightforward manner in UV complete particle models. However, we would like instead to extend as little as possible a simplified model consisting of the DM particle and its mediator, the elements needed for tree-level calculations,  to preserve its generality. 
Several calculations exist in this spirit  of loop-level 
corrections to the interactions mediated by a pseudoscalar particle~\cite{Arcadi:2017wqi,Bell:2018zra,Li:2018qip,Abe:2018emu,Ertas:2019dew}. This type of interaction is particularly favorable to produce sizable loop-level contributions to the interaction rate at low recoil energies because at tree-level the cross section is momentum transfer suppressed  (see e.g.~table 1 and eqs.~(2.31) and~(2.33) of ref.~\cite{Gelmini:2018ogy}). 
Using these loop-level calculations, here we compare the argon and xenon-based direct detection capabilities of fermionic DM with interactions mediated by a pseudoscalar boson with tree-level couplings $\bar{f} \gamma_5 f$ with Standard Model (SM) fermions $f$.

The paper is structured as follows. In section~\ref{sec:model} we discuss the general interaction model for the DM particle and the SM fermions. In section~\ref{sec:loop} we discuss the loop-level corrections to the interactions. In section~\ref{sec:directdet} we present the expressions used for the computation of direct detection event rates. In section~\ref{sec:comparison} we present our results on the detection capability of Ar and Xe-based detectors, and we conclude in section~\ref{sec:conclusions}.

\section{The interaction model}
\label{sec:model}

Models of DM particles with a pseudoscalar mediator have been studied 
for more than a decade~\cite{Freytsis:2010ne,Dienes:2013xya,Boehm:2014hva}, and
UV complete models including interaction of this type have been studied at colliders (see e.g.~\cite{Ipek:2014gua,No:2015xqa,Goncalves:2016iyg,Bauer:2017ota,Bauer:2017fsw,LHCDarkMatterWorkingGroup:2018ufk}). 

Loop-level contributions to a cross section depend on all the particles that can be exchanged in a particular theory, not only on the mediator which dominates the interaction at the tree-level. Thus, without postulating a complete model, it is of particular importance at this level to include the necessary couplings of the pseudoscalar mediator to the SM Higgs boson  required by gauge invariance~\cite{Arcadi:2017wqi,Bell:2018zra,Li:2018qip,Abe:2018emu, Ertas:2019dew}. When these are included, even working at one-loop level, two-loop diagrams cannot be neglected~\cite{Abe:2018emu, Ertas:2019dew}. 

 Following ref.~\cite{Ertas:2019dew}, we assume here a simple model for the  interaction of a Dirac fermion DM particle $\chi$ and SM fermions $f$ mediated by a spin-zero boson field $\phi$ with mass $m_\phi$. Two different types of Lagrangian terms  are assumed: S-PS and PS-PS. Here, the first S (scalar) or PS (pseudoscalar) refers to the DM coupling without or with a $\gamma_5$, respectively,  and the second PS to the SM fermion coupling with a $\gamma_5$. These are respectively
\begin{align}
\mathcal{L_{\rm S-PS}} = g_\chi\,\phi\,\bar{\chi}\, \chi + g_\text{SM}\sum_{f} \frac{m_f}{\varv} \phi\,\bar{f}\, i \gamma_5 \,f\, ,
\label{eq:S-PS}
\end{align}
and
\begin{align}
\mathcal{L_{\rm PS-PS}} = g_\chi\,\phi\,\bar{\chi}\,  i \gamma_5 \, \chi + g_\text{SM}\sum_{f} \frac{m_f}{\varv} \phi\,\bar{f}\, i \gamma_5 \,f\, ,
 \label{eq:PS-PS}
\end{align}
where $\varv= 246\,\mathrm{GeV}$ is the electroweak vacuum expectation value,  $g_\chi$ and  $(g_\text{SM}~{m_f}/{\varv})$ denote the coupling constants of the mediator $\phi$ to the DM and SM fermions, respectively, and $m_f$ are the SM fermion masses. The couplings of the new field $\phi$ to SM fermions are taken to be proportional to the  Yukawa-couplings $({m_f}/{\varv})$, in agreement with the hypothesis of minimal flavour violation~\cite{DAmbrosio:2002vsn}, which leads to weakened  flavor physics constraints on them.

Gauge symmetry requires that $\phi$ couples to the SM Higgs field in a quartic term in the scalar potential. After the electroweak spontaneous symmetry breaking, this quartic coupling yields the term
\begin{align}
\mathcal{L}^\text{Higgs}_\text{int} = \frac{1}{2} \lambda_{\phi h} \varv h \phi^2\;,
\end{align}
where $h$ is the Higgs boson field. This term is identified in ref.~\cite{Ertas:2019dew} as the most relevant  $\phi$-$h$ coupling, because other interaction terms involving two Higgs bosons do not give any relevant contribution to the calculation of direct detection signatures.   The  constant $\lambda_{\phi h}$  is  an additional free parameter of the model, constrained to be  $\lambda_{\phi h} \lesssim 0.01$ when the invisible decays $h \to \phi \phi$ are kinematically possible, i.e. when $m_\phi < m_h/2$ (see figure~6 of ref.~\cite{Ertas:2019dew}).

The parameters of this simple model are, therefore, five:  the mass of the DM particle $m_\chi$,  the mass of the mediator $m_\phi$, 
and the three coupling constants $g_\chi,  g_\text{SM}$ and $\lambda_{\phi h}$. As explained below, in the following $m_\phi$ is taken to be smaller than the t quark mass and larger than the b quark mass, $m_b < m_\phi \ll m_t$, for consistency of the calculations~\cite{Ertas:2019dew}.

\section{Loop-level corrections}
\label{sec:loop}

The two initial interactions in eqs.~\eqref{eq:S-PS} and~\eqref{eq:PS-PS}, given in the first column of table~\ref{tab:1}, 
generate at the loop-level an effective low energy Lagrangian which contains terms proportional to some of the other possible interactions~\cite{Ertas:2019dew}, indicated by an $X$ in the table. S-S refers to couplings $[\bar{\chi}\chi ~ \bar{f} f]$ and PS-S to  $[\bar{\chi}\gamma_5\chi ~ \bar{f} f]$,   which are both independent of the nuclear spin. Namely, at the loop-level 
S-PS  interactions generate S-S, S-PS and PS-PS terms, while starting from PS-PS interactions, terms proportional to all four interaction types are produced at the loop-level.  

\begin{table}[t]
 \begin{center}
  \begin{tabular}{ | l | c |c| c | c |}
    \hline
    Initial & Effective  S-S & Effective PS-S  & Effective S-PS & Effective PS-PS \\ \hline
    S-PS & X & 0 & X & X \, \\ \hline
    PS-PS & X & X & X & X \, \\
    \hline
  \end{tabular}
 \caption{The two initial couplings in eqs.~\eqref{eq:S-PS} and \eqref{eq:PS-PS}, shown in the first column, 
generate at the loop-level an effective low energy Lagrangian with terms proportional to some of the other interactions indicated by an $X$~\cite{Ertas:2019dew}. S-S refers to couplings $(\bar{\chi}\, \chi \bar{f}\, f)$ and PS-S to  $(\bar{\chi}\, \gamma_5 \, \chi \bar{f}\, f)$,   which are both independent of the nuclear spin. }
\label{tab:1}
\end{center}
\end{table}

We utilize the effective approach developed in  ref.~\cite{Ertas:2019dew} for calculations of two-loop processes contributing to the effective DM-gluon interactions,  which can replace the full two-loop calculations. In this approach, the two-loop processes can be decomposed into two one-loop diagrams by first integrating out the top quark, and then the mediator $\phi$. In order to apply this approach, we need to take $m_\phi \ll m_t$. In this case,  
ref.~\cite{Ertas:2019dew} shows that their  effective approach for the two-loop calculation agrees very well with the full two-loop calculations for all values of $m_\chi$ (see figure~4 of ref.~\cite{Ertas:2019dew}), and also
that  when $m_\phi> m_b$ the contribution from bottom and charm quarks can be neglected.\footnote{Loop processes for $m_\phi \gg m_t$, and more general electroweak structures, were calculated in ref.~\cite{Bishara:2018vix}.}

After matching the effective DM-quark and effective DM-gluon interactions onto non-relativistic DM-nucleon interactions, the effective Lagrangian can be written as~\cite{Ertas:2019dew},
\begin{equation}
\mathcal{L}_{\chi N}^{\rm eff}=\left(\mathcal{C}_{{\rm eff}, N}^{\rm SI}\bar{\chi}\chi + \mathcal{C}_{{\rm eff}, N}^{\rm SI, CPV}\bar{\chi} i \gamma_5\chi\right) \bar{N}N+\left(\mathcal{C}_{{\rm eff}, N}^{\rm SD, CPV}\bar{\chi}\chi + \mathcal{C}_{{\rm eff}, N}^{\rm SD}\bar{\chi} i \gamma_5\chi\right) \bar{N}i\gamma_5 N,
\label{eq:effL}
\end{equation}
where $N=p,n$ indicates proton or neutron fields. The coefficients $\mathcal{C}_{{\rm eff}, N}$ are given in ref.~\cite{Ertas:2019dew} and depend on tree-level and loop-level coefficients. The labels ``SI" and ``SD" refer to nuclear-spin dependent and nuclear-spin independent terms, and the label ``CPV" refers to CP violation and is used for coefficients that exist only when CP is violated. 

We refer the reader to ref.~\cite{Ertas:2019dew} for detailed expressions of the $\mathcal{C}_{{\rm eff}, N}$ coefficients, and only mention here their dependence on the Passarino-Veltman functions $C_0$ and $C_2$~\cite{Passarino:1978jh}, as well as the $X_2$, $Y_2$, and $Z$ functions defined in ref.~\cite{Abe:2015rja},
\begin{align}
\mathcal{C}_{{\rm eff}, N}^{\rm SI}&=\mathcal{C}_{{\rm eff}, N}^{\rm SI}(\mathcal{C}^{\rm tree},C_0, C_2, X_2, Y_2, Z_{00}, Z_{001}, Z_{111}),\nonumber\\  
\mathcal{C}_{{\rm eff}, N}^{\rm SI, CPV}&=\mathcal{C}_{{\rm eff}, N}^{\rm SI, CPV}(\mathcal{C}^{\rm tree}, C_0, X_2, Z_{00}, Z_{11}),\nonumber\\ 
\mathcal{C}_{{\rm eff}, N}^{\rm SD}&=\mathcal{C}_{{\rm eff}, N}^{\rm SD}(\mathcal{C}^{\rm tree},C_0, X_2),\nonumber\\ 
\mathcal{C}_{{\rm eff}, N}^{\rm SD, CPV}&=\mathcal{C}_{{\rm eff}, N}^{\rm SD, CPV}(\mathcal{C}^{\rm tree}, C_0, C_2, X_2, Y_2),
\end{align}
where
\begin{equation}
\mathcal{C}^{\rm tree} = \frac{g_\chi g_{\rm SM}}{\varv m_\phi^2}.
\end{equation}
For the on-shell condition $p^2 = m_\chi^2$, where $p$ is the DM particle momentum, the functions $C_0$, $C_2$, $X_2$, $Y_2$, as well as the $Z$ functions,  only depend on $m_\chi$ and $m_\phi$, and can be calculated using \texttt{Package-X}~\cite{Patel:2015tea, Patel:2016fam} and \texttt{CollierLink}~\cite{Denner:2016kdg} (see also appendix A of ref.~\cite{Ertas:2019dew}). Notice that the $\mathcal{C}_{{\rm eff}, N}$ coefficients also depend on the nuclear form factors for the quark and gluon contents of the nucleon. For the SD coefficients, the nuclear form factors  depend on the momentum  $q=|\vec{q}|$ exchanged between the DM particle and the nucleon, due to contributions from the $\pi$ and $\eta$ mesons~\cite{Bishara:2017pfq}.\footnote{ Updated nuclear form factors have been very recently given in ref.~\cite{Haxton:2024lyc}.}

In general, the SI coefficients $\mathcal{C}_{{\rm eff}, N}^{\rm SI}$ and $\mathcal{C}_{{\rm eff}, N}^{\rm SI, CPV}$ depend on the five parameters $m_\chi$, $m_\phi$, $g_\chi$, $g_{\rm SM}$, and $\lambda_{\phi h}$.
The SD coefficients $\mathcal{C}_{{\rm eff}, N}^{\rm SD}$ and $\mathcal{C}_{{\rm eff}, N}^{\rm SD, CPV}$ depend on those same parameters in addition to the momentum transfer $q$. 

The effective Lagrangian in eq.~\eqref{eq:effL} can be parametrized in terms of a set of effective operators in the non-relativistic limit~\cite{Fan:2010gt, Fitzpatrick:2012ix, Fitzpatrick:2012ib, Anand:2013yka, Dent:2015zpa},
\begin{equation}
\mathcal{L}_{\chi N}^{\rm eff} \to \sum_i c_i^N \mathcal{O}_i^N.
\end{equation}
For the model we are considering, only the following four effective operators contribute,
\begin{align}
\label{eq:operators}
\mathcal{O}_1^N &=1_\chi 1_N,\nonumber\\
\mathcal{O}_6^N &=\left(\vec S_\chi \cdot \frac{\vec q}{m_N} \right) \left(\vec S_N \cdot \frac{\vec q}{m_N} \right),\nonumber\\
\mathcal{O}_{10}^N &=i\, \vec S_N \cdot \frac{\vec q}{m_N},\nonumber\\
\mathcal{O}_{11}^N &=i\, \vec S_\chi \cdot \frac{\vec q}{m_N},
\end{align}
where $\vec S_\chi$ is the DM particle spin, $\vec S_N$ is the nucleon spin, and $m_N$ is the nucleon mass. The coefficients corresponding to these operators are respectively $c_1^N=\mathcal{C}_{{\rm eff}, N}^{\rm SI}$, $c_6^N=(m_N/m_\chi)~\mathcal{C}_{{\rm eff}, N}^{\rm SD}$, $c_{10}^N=\mathcal{C}_{{\rm eff}, N}^{\rm SD, CPV}$, and $c_{11}^N=-(m_N/m_\chi)~\mathcal{C}_{{\rm eff}, N}^{\rm SI, CPV}$.

\section{Direct detection rates}
\label{sec:directdet}

With the operators given in eq.~\eqref{eq:operators}, we calculate the spin averaged DM-nuclide, $T$, transition probability~\cite{Anand:2013yka},
\begin{equation}
    P_{{\rm tot}, T}=\frac{1}{2j_\chi+1} \frac{1}{2j_T+1}\sum_{\rm spins}|\mathcal{M}_T|^2~,
\end{equation}
where $j_\chi$ and $j_T$ are the total angular momentum of the DM and the target nucleus respectively and $\mathcal{M}_T$ is the DM-nucleus scattering amplitude summed over nucleons and computed between nuclear states (defined e.g.~in eq.~(40) of ref.~\cite{Anand:2013yka}). This transition probability is then used to calculate the DM-nucleus differential cross-section,
\begin{equation}
    \frac{d\sigma_T}{d E_R}=\frac{m_{T}}{2 \pi v^2}P_{{\rm tot}, T}\,
\end{equation}
 where  $E_R$ denotes the recoil energy of the target nucleus, $m_T$ is the mass of the target nucleus, and $\vec v$ is the relative velocity between the DM and the target nucleus, with $v\equiv |\vec v|$. Throughout our calculations, we assume that there is no difference between the recoil energy and the observed energy; therefore, we do not consider an experimental energy resolution. 

In the case that the detector includes multiple  nuclides, the differential event rate per  unit energy, detector mass, and time is given by 
\begin{equation}
    \frac{dR}{dE_R}=\sum_T\frac{dR_T}{dE_R}=\sum_T \frac{C_T}{m_T}\,\frac{\rho_\chi}{m_\chi}\int_{v\geq v_{{\rm min}, T}} d^3 v\, \frac{d\sigma_T}{dE_R}\,v\, f(\vec{v}, t)~,
\end{equation}
where the sum is over the different isotopes or target nuclides, $T$, present in the detector, $C_T$ is the mass fraction of $T$ in the detector, $\rho_\chi$ is the local DM density, and $f(\vec v, t)$ is the local DM velocity distribution in the Earth's rest frame. The minimum DM speed to produce a recoil energy $E_R$ in the detector is
\begin{equation}
    v_{{\rm min}, T}=\frac{q}{2\mu_T}=\sqrt{\frac{m_TE_R}{2\mu_T^2}}~,
\end{equation}
where $\mu_T$ is the DM-target nucleus reduced mass.

We assume the Standard Halo Model~\cite{Drukier:1986tm} for describing the local distribution of DM in the galactic halo. Recent cosmological simulations including both DM and baryons show that the  Maxwellian distribution fits well the local DM velocity distribution of Milky Way-like galaxies~\cite{Bozorgnia:2016ogo, Kelso:2016qqj, Sloane:2016kyi, Poole-McKenzie:2020dbo}. However, DM from massive satellites such as the Large Magellanic Cloud can significantly  impact the high speed tail of the local DM velocity distribution~\cite{Reynoso-Cordova:2024xqz, Smith-Orlik:2023kyl, Besla:2019xbx, Donaldson:2021byu}. For simplicity, we neglect these effects in this work. 

The local DM density is set to $\rho_\chi=0.3$~GeV$/$cm$^3$. The local DM velocity distribution is modelled as a Maxwell-Boltzmann distribution in the galactic rest frame, and truncated at the local escape speed, $v_{\rm esc}$, from the galaxy,
\begin{equation}
    f_{\rm gal}(\vec{v})=\frac{1}{N_{\rm esc} (v_0 \sqrt{\pi})^3} \left[ e^{-v^2/v_0^2}-e^{-v_{\rm esc}^2/v_0^2}\right]
\Theta (v_{\rm esc}^2-\vec{v}^2),
\label{eq:fv}
\end{equation}
where
\begin{equation}
   N_{\rm esc}=\erf\left(v_{\rm esc}/v_0 \right)-\frac{2}{\sqrt{\pi}}\left(\frac{v_{\rm esc}}{{v_0}}\right) e^{-v_{\rm esc}^2/v_0^2}\left[1+\frac{2}{3}\left(\frac{v_{\rm esc}}{v_0}\right)^2\right].
\end{equation}
Here, $v_0$ is the local circular speed, and the velocity distribution is normalized such that $\int d^3 v f_{\rm gal}(\vec v)=1$. The second term in the bracket in eq.~\eqref{eq:fv} ensures that the velocity distribution falls off smoothly to zero at $v_{\rm esc}$. We take $v_0= 235$~km/s based on measurements of galactic masers~\cite{Reid:2009nj, Bovy:2009dr}, and $v_{\rm esc}=550$~km/s based on measurements of high velocity stars from the RAVE survey~\cite{Smith:2006ym}.

We transform the DM velocity distribution from the galactic rest frame to the Earth's rest frame, 
\begin{equation}
 f(\vec v, t)=f_{\rm gal}\left(\vec v + \vec v_s+ \vec v_e(t)\right)\, ,
\end{equation}
where $\vec v_s=\vec v_c+\vec v_{\rm pec}$ is the velocity of the Sun in the Galactic rest frame, and $\vec v_e(t)$ is  Earth's velocity with respect to the Sun, which we neglect in this work. Here, $\vec v_c$ is the Sun's circular velocity (where we take $v_c=v_0=235$~km/s), and $\vec v_{\rm pec}=(11.10,12.24, 7.25)$~km/s~\cite{Schoenrich:2009bx} is the peculiar velocity of the Sun with respect to the Local Standard of Rest, in galactic coordinates. 

\section{Comparison of Ar and Xe detection capability}
\label{sec:comparison}

\begin{figure}[tb]
    \centering
    \includegraphics[width=0.47\textwidth]{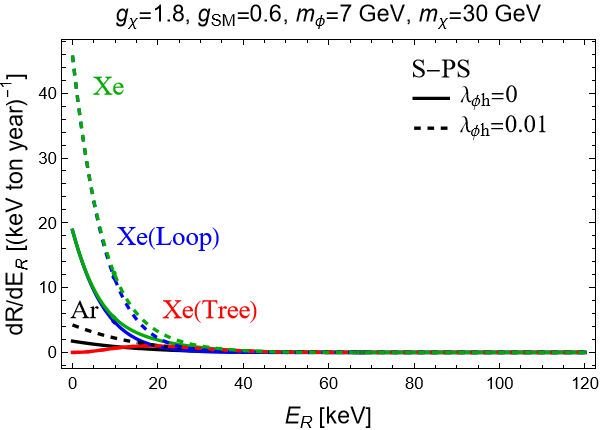}
    \hspace{10pt}\includegraphics[width=0.47\textwidth]{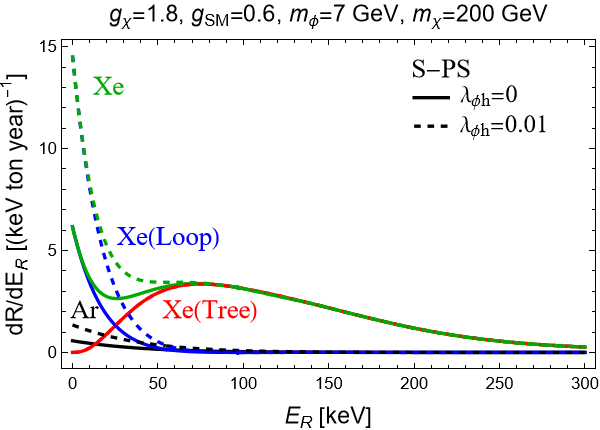} 
    \includegraphics[width=0.49\textwidth]{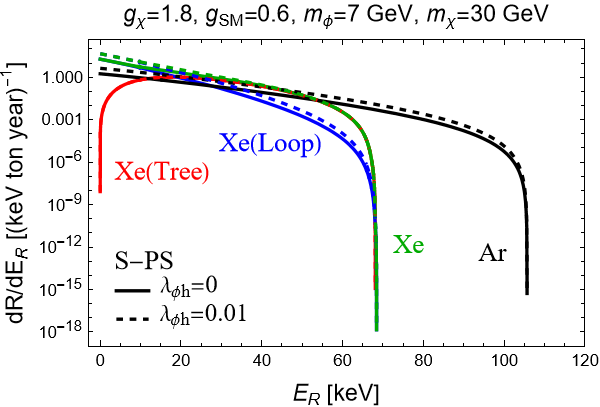}
    \includegraphics[width=0.49\textwidth]{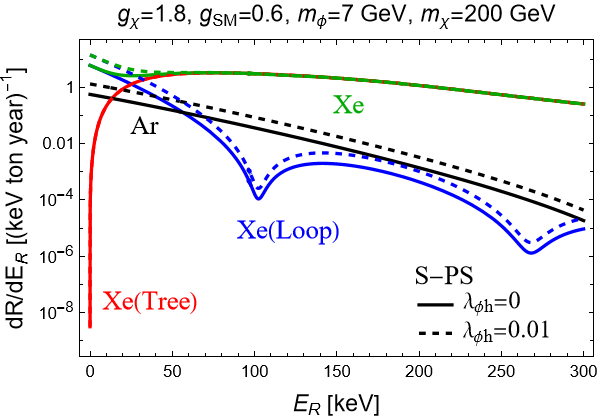}
                \caption{Differential event rates for S-PS Lagrangian couplings, with $g_\chi=1.8$, $g_{\rm SM}=0.6$,  $m_\phi=7$~GeV, $m_\chi=30$~GeV (left panels) and $m_\chi=200$~GeV (right panels), $\lambda_{\phi h}=0$ (solid lines) and $\lambda_{\phi h}=0.01$ (dashed lines) for Ar (black lines) and Xe (green lines) targets. The contributions from tree-level only (red lines) and loop-level only (blue lines) in Xe are also shown. The rates are shown in linear scale in the upper panels and  in logarithmic scale in the lower panels.}
    \label{fig:dRdEphix0}
\end{figure}

\begin{figure}[tb]
    \centering
    \includegraphics[width=0.47\textwidth]{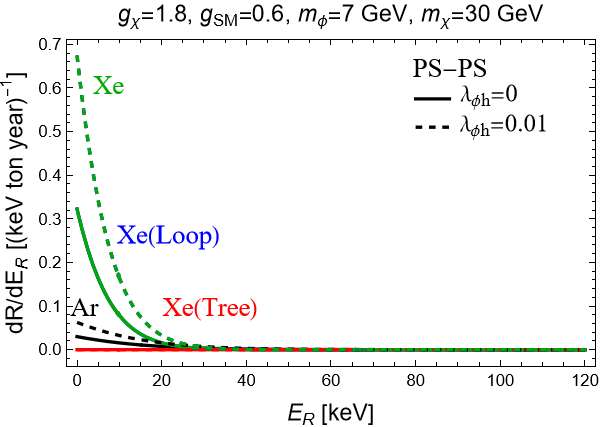}
    \includegraphics[width=0.49\textwidth]{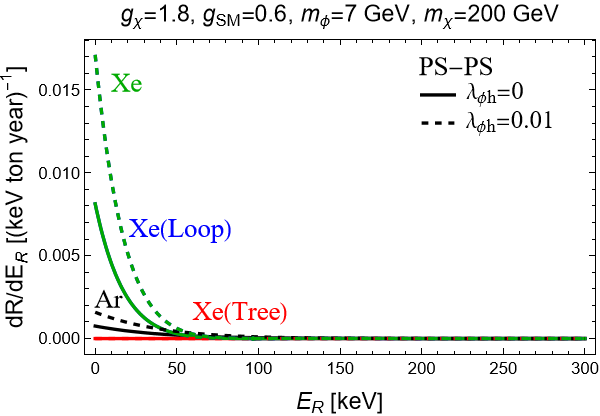} 
    \includegraphics[width=0.49\textwidth]{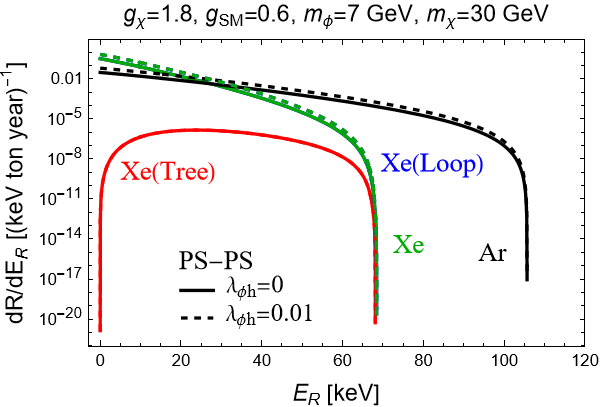}
    \includegraphics[width=0.49\textwidth]{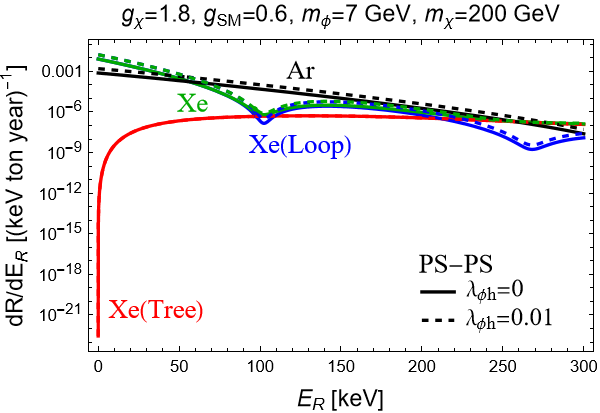}
    \caption{Same as figure~\ref{fig:dRdEphix0} but for PS-PS Lagrangian couplings. Notice that the rates are roughly two orders of magnitude smaller than those in figure~\ref{fig:dRdEphix0}. The loop level contributions for Xe (blue lines) follow closely the total rate (green lines) and are therefore not clearly visible in the two top and the bottom left panels.
    }  
    \label{fig:dRdEphixpi1}
\end{figure}

Figures~\ref{fig:dRdEphix0} and \ref{fig:dRdEphixpi1}  
show the predicted differential event rates for S-PS and PS-PS Lagrangian couplings, respectively, in Ar and Xe targets in linear and logarithmic scales.  These figures show  the tree-level only (red lines),  loop-level contributions only (blue lines) and the total rate (green lines) in Xe,  and the total rate entirely due to loop-level contributions (black lines) in Ar targets, for some values of the relevant parameters chosen as examples, i.e. $m_\phi= 7$~GeV, $g_\chi=1.8$, $g_\text{SM}=0.6$, the minimum and the maximum possible values of the $\phi$-$h$ coupling, $\lambda_{\phi h}=0$ (solid lines) and $\lambda_{\phi h}= 0.01$ (dashed lines), and two values of the DM particle mass,  $m_\chi=30$~GeV and $m_\chi=200$~GeV. The  event rates have been computed using the Mathematica package \texttt{dmformfactor}~\cite{Anand:2013yka}\footnote{Updated shell model one body density matrices for Ar and Xe are now available in the very recent open-source code \href{https://github.com/Berkeley-Electroweak-Physics}{\texttt{MuonBridge}}~\cite{Haxton:2024lyc}.}.

Figures~\ref{fig:dRdEphix0} and~\ref{fig:dRdEphixpi1}   clearly show that the  loop-level contributions are important for low recoil energies, and that the 
expected rates for S-PS Lagrangian couplings are much larger, by a factor of about $10^2$,  than those for PS-PS Lagrangian couplings.  In both cases, the rates increase with $\lambda_{\phi h}$ by a factor of order 1 when going from the minimum to the maximum of this coupling. Also in both cases, even at the lowest recoil energies,  the rate expected in an Ar target is smaller by about one order of magnitude than in a Xe target.

\begin{figure}
    \centering
    \includegraphics[width=0.49\textwidth]{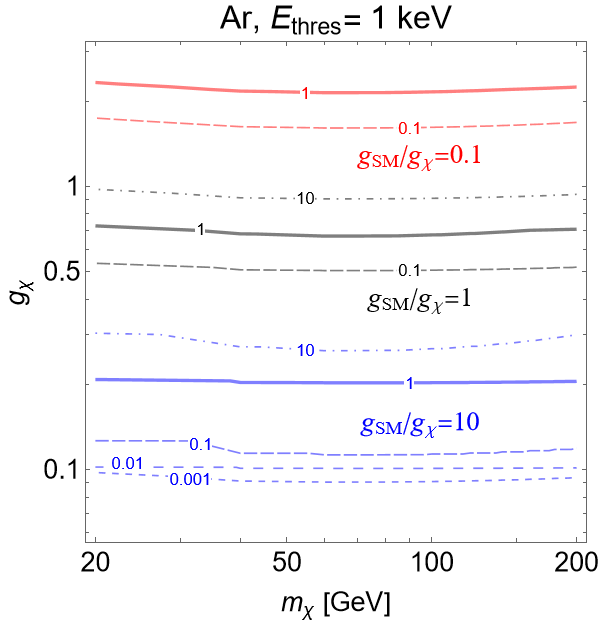}
    \includegraphics[width=0.49\textwidth]{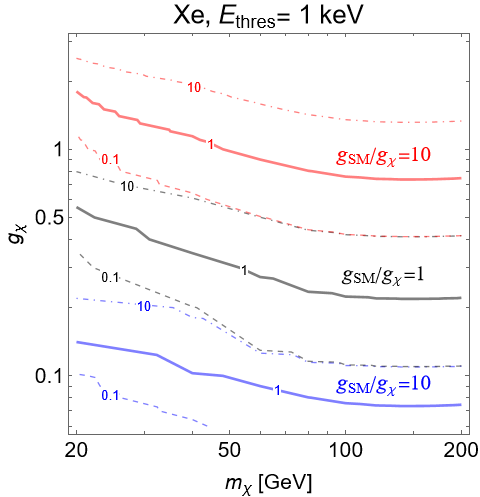}
    \caption{Lines of equal energy-integrated event rate for $E_R > 1$ keV in the $m_\chi$-$g_\chi$ plane 
    for Ar (left panel) and Xe (right panel) targets for a model with S-PS Lagrangian couplings, $m_\phi= 7$~GeV and $\lambda_{ah}=0$, for three different values of the $g_{\rm SM}/g_\chi$ coupling ratios, 0.1 (red lines),  1 (black lines) and 10 (blue lines). For each of these ratios different  lines are shown corresponding to an integrated rate shown in the respective labels in units of (ton year)$^{-1}$, e.g. 10 (dot-dashed lines above the solid lines), 1 (solid lines) and 0.1 (dashed lines below the solid lines).
    }
    \label{fig:Ar-Xe-Events}
\end{figure}

\begin{figure}
    \centering
    \includegraphics[width=0.49\textwidth]{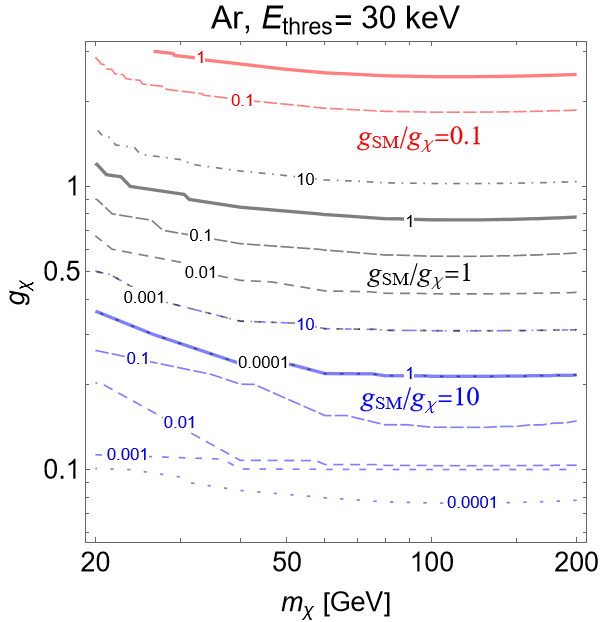}
    \caption{Same as figure~\ref{fig:Ar-Xe-Events}, but only for Ar and assuming an energy threshold of $E_R= 30$ keV instead of 1 keV. Notice that the solid blue line corresponding to an integrated rate of 1/(ton year) for $g_{\rm SM}/g_\chi=10$ is overlayed on the dotted black line corresponding to 0.0001/(ton year) for $g_{\rm SM}/g_\chi=1$.
     }
        \label{fig:Ar30-Events}
\end{figure}

To compare the detection capabilities of Ar and Xe-based detectors, we first consider idealized experimental conditions in which we ignore experimental backgrounds and efficiencies, and compute the event rates in each detector. Because loop-level contributions are only important at low energies,  the energy threshold of future detectors will be very important for detecting rates enhanced by them.  
Although both argon and xenon-based detectors may achieve nuclear recoil energy thresholds below 1 keV using their S2 (ionization) signal only~\cite{DarkSide:2018bpj, DarkSide-50:2023fcw, Aalbers:2022dzr}, in figure~\ref{fig:Ar-Xe-Events} we show lines of equal energy-integrated rates taking a threshold of 1 keV for both. This is the targeted threshold for future xenon-based detectors  such as DARWIN/XLZD~\cite{Baudis:2024jnk,aalbers2024xlzd} and for argon-based detectors such as DarkSide-LowMass (DS-LM) for ~\cite{Wada:2024xqq}.  We also consider the possibility of Ar-based detectors with a larger threshold, such as the 30 keV expected for DarkSide-20k~\cite{DarkSide-20k:2017zyg} and show the corresponding lines of equal integrated rate in figure~\ref{fig:Ar30-Events}.
Figure~\ref{fig:Ar-Xe-Events} shows lines of equal integrated rate above the threshold of $E_R= 1$~keV  in the $m_\chi$-$g_\chi$ plane, for Ar (left panel) and Xe (right panel) targets, for the most favorable type of Lagrangian coupling of the two we considered, the S-PS.  The mediator mass is chosen to be $m_\phi= 7$~GeV, as in previous figures, and $\lambda_{ah}=0$, which is one of the two values in previous figures.  Lines corresponding to an integrated rate of 10 (dot-dashed lines above the solid lines), 1 (solid lines) and 0.1 (dashed lines below the solid lines) in units of (ton year)$^{-1}$ are given for 
three different values of the $g_{\rm SM}/g_\chi$ coupling ratios, 0.1 (in red),  1 (in black) and 10 (in blue). A few more blue lines corresponding to lower values of integrated rates, as labeled, are given for argon. Notice that we can integrate  the total differential event rates (solid lines) in figure~\ref{fig:dRdEphix0} above 1~keV to approximately obtain the value close to the red lines in figure~\ref{fig:Ar-Xe-Events} for $m_\chi=30$~GeV and 200 GeV, $g_\chi= 1.8$ and $g_{\rm SM}/g_\chi \simeq 0.3$. 

Figure~\ref{fig:Ar-Xe-Events} allows us to estimate the difference in target mass between idealized argon- and xenon-based detectors that would be required to detect a similar number of events per year, if both have an energy threshold of 1 keV, for particular values of the parameters, assuming the same mediator mass as in the previous figures and $m_\chi$ in the 20 to 200 GeV range. For example, for a 30 GeV DM particle mass,  $g_\chi= 0.2$, and $g_{\rm SM}=2$, the solid blue line in the left panel  shows that an integrated rate of 1$/$(ton year) is expected in Ar and the right panel the dot-dashed blue line indicates that 10$/$(ton year)  would be expected in Xe. Thus, to have a similar rate in both targets (ignoring backgrounds and efficiencies) one would require a 10 times larger mass in Ar than in Xe.   The difference becomes more accentuated for larger DM masses, e.g.~for a DM mass of 200~GeV, the rate in argon for the same coupling would still be close to 1$/$(ton year), while in Xe the expected rate is  $\gg 10/$(ton year) (curve not shown in the figure).

Figure~\ref{fig:Ar30-Events} shows lines of equal integrated rate for an Ar target in the $m_\chi$-$g_\chi$ plane for the same couplings and parameters used in figure~\ref{fig:Ar-Xe-Events} (i.e.~for  the S-PS Lagrangian coupling, with $m_\phi= 7$~GeV and $\lambda_{ah}=0$), but  above the threshold of $E_R= 30$ keV instead of 1 keV.  This figure clearly  shows the disadvantage that a larger energy threshold of 30 keV  would have to detect DM particle with S-PS couplings in an  Ar-based detector.  This would make it practically impossible for Ar to detect this type of interaction. For example, for a 30 GeV DM particle,  $g_\chi= 0.2$, and $g_{\rm SM}=2$, the medium-dashed blue line in figure~\ref{fig:Ar30-Events} shows that an integrated rate of between 0.01$/$(ton year) and 0.1$/$(ton year) is expected in Ar. This is more than an order of magnitude smaller than the rate expected in Ar with an energy threshold of 1 keV for the same couplings. 

We do not present specific figures for the integrated rates in the PS-PS case. Our general conclusions comparing both targets would be similar to those for S-PS, but the event rates would be around two orders of magnitude smaller with respect to S-PS, requiring an experimental exposure of $\sim 100$ times larger to have a signal in the PS-PS case.

\begin{figure}[tb]
    \centering
    \includegraphics[width=0.49\textwidth]{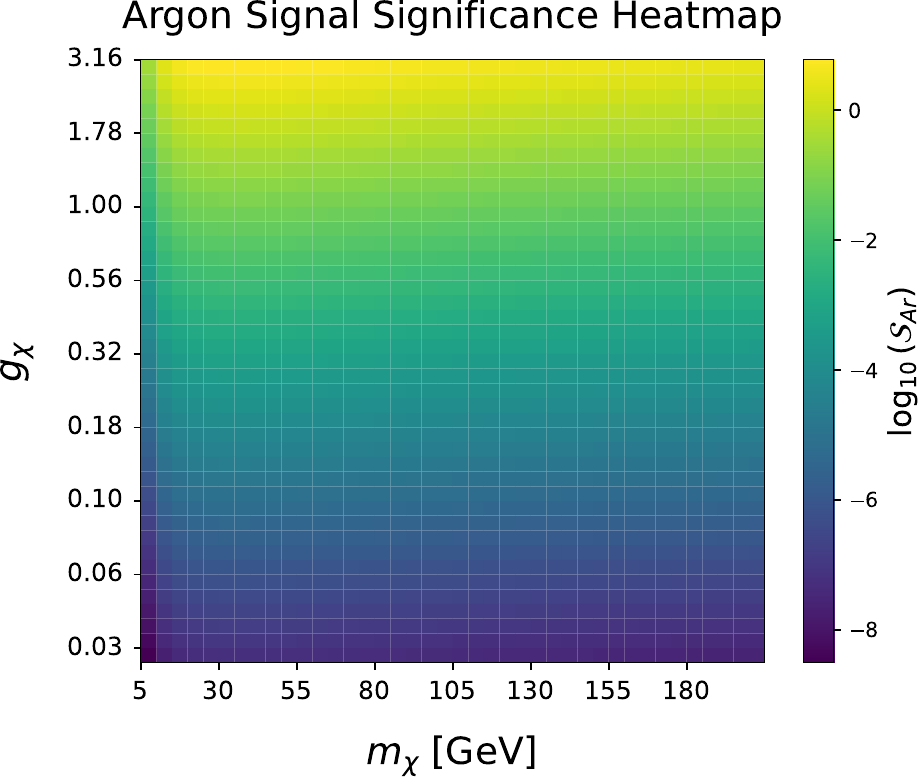}
    \includegraphics[width=0.49\textwidth]{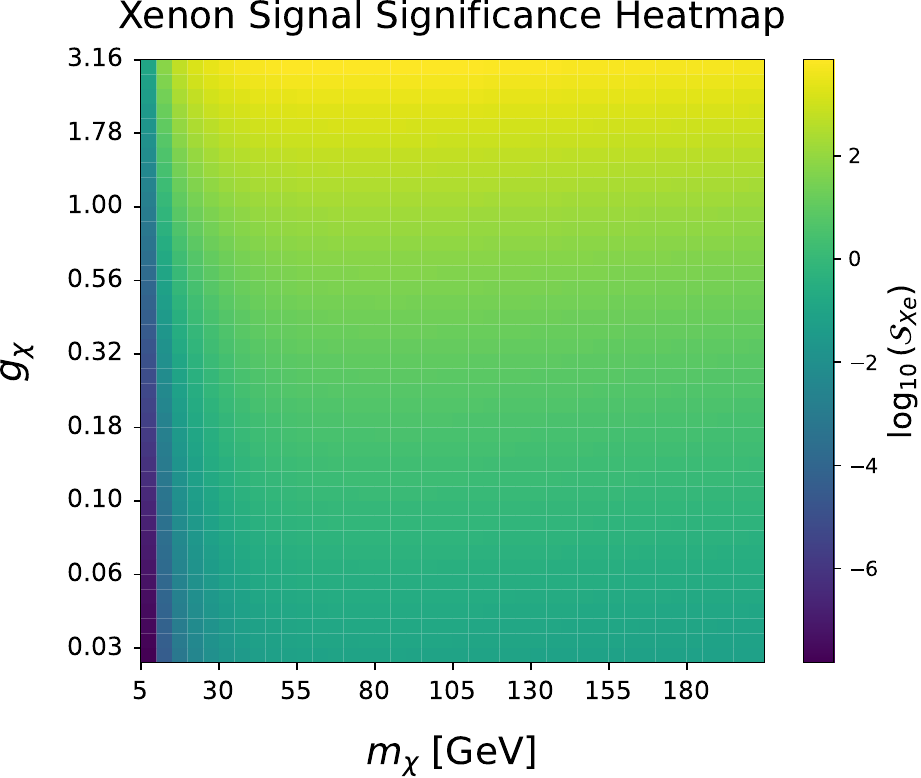} 
    \caption{The signal significance, $\mathcal{S}$, heat map for the proposed  DS-LM argon experiment (left) and the LZ xenon experiment (right).
    The model parameters assumed are $m_\phi = 7$ GeV, $\lambda_{\phi h} =0.01$ and $g_\text{SM} =0.6$ for S-PS Lagrangian couplings. As expected, in both target materials, the signal significances are larger at large DM couplings, indicating a better detection probability.
    The exposures being considered are 1 ton year for DS-LM~\cite{GlobalArgonDarkMatter:2022ppc} and 15.3 ton year for LZ~\cite{Mount:2017qzi}.
    $\mathcal{S}$ in Ar grows faster with $g_\chi$, as the entire rate
    comes from loop-level contributions with Lagrangian coefficients scaling as $g_\chi^2$.
    It also drops slower at smaller DM masses due to the fact that Ar is lighter than Xe. 
    }  
    \label{fig:heatmaps}
\end{figure}

Next, we compare the detection capabilities of Ar- and Xe-based detectors in the context of realistic experimental conditions. For the Ar experiment, we choose the DS-LM  proposal of the GADMC~\cite{GlobalArgonDarkMatter:2022ppc} for its low background level, in addition to its low energy threshold. Since the S1 pulse-shape electron-nuclear recoil type discrimination requires  an energy deposit of about 30 keV to work~\cite{Fan:2016ymy,Cadeddu:2018osv} --which is larger than typical energy deposits 
from loop-level rate contributions-- we utilize the S2-only channel in this analysis, with a threshold of four electrons (4$e^-$). Here, we use the detector response model established in refs.~\cite{DarkSide:2021bnz,GlobalArgonDarkMatter:2022ppc} to convert the true recoil energy spectrum into the number of electrons, and assume a total exposure of 1 ton year~\cite{GlobalArgonDarkMatter:2022ppc}. We adopt the background model from ref.~\cite{GlobalArgonDarkMatter:2022ppc}, with three dominant background components considered: the detector $\gamma$ scatters, $^{39}$Ar $\beta$ decays, and the solar neutrino coherent scatters.
As the cut acceptance loss due to cuts other than the fiducialization cut is typically negligible for S2's larger than a few electrons \cite{DarkSide:2018bpj}, we do not further consider the acceptance effect of those cuts throughout this analysis.
We quantify the detector's sensitivity to certain model parameters using the signal significance $\mathcal{S}=N_S/\sqrt{N_B}$, where $N_S$ and $N_B$ are the expected signal and background events, respectively.
The significance heat map of models in the $m_\chi-g_\chi$ plane  for S-PS interaction type with $m_\phi = 7$ GeV, $\lambda_{\phi h} =0.01$, $g_\text{SM} =0.06$, and $m_\chi$ in the 5 to 205 GeV range is shown in the left panel of figure~\ref{fig:heatmaps}.

For the Xe-based experiment, we use the LZ experiment \cite{Mount:2017qzi} as a benchmark due to its leading sensitivity to nuclear recoil signatures and low background rate, with the detector configuration specified as in its first science run results~\cite{LZ:2022lsv}.
We use \texttt{NEST v2.3.11}~\cite{Szydagis:2011tk} to simulate the detector response, with parameters adjusted to the calibration data in \cite{LZ:2022lsv}.
\texttt{NEST} simulated events in the $S_1, S_2$ space are then converted into the reconstructed energy using the relation $E_\text{rec} = W_q(S_{1}/g_1 + S_{2}/g_2)$, where $W_q = 13.7$ eV is the Xe work function, $g_1 = 0.114$ is the photon collection efficiency, and $g_2 = 47.1$ is the ionization gain~\cite{LZ:2022lsv}.
The background model in the reconstructed energy space and cut acceptances for nuclear recoil signals are also adopted from ref.~\cite{LZ:2022lsv}.
Most of the background components are electron-recoil-like, and Xe-based detectors are capable of discriminating between electron-recoils and nuclear-recoils at lower energy using S1/S2 ratio.
This would further enhance the sensitivity of Xe-based detectors to low-energy nuclear-recoil signals for LZ. We conservatively assume a flat 99.5$\%$ electron-recoil rejection efficiency \cite{Mount:2017qzi}.
A total exposure of 15.3 ton year~\cite{Mount:2017qzi} is used to calculate the signal significance, $\mathcal{S}$, heat map  for LZ. 
The result for the same model parameters we used for argon can be found in the right panel of figure~\ref{fig:heatmaps}.

From figure~\ref{fig:heatmaps} we can see that for both the DS-LM argon and LZ xenon detectors, the largest signal significance is obtained for large $g_\chi$ values. Comparison of the two panels of the figure shows that for most of the parameter space, the signal significance for LZ is larger than that of DS-LM by a factor ranging from an order of magnitude to up to several orders of magnitude depending on $m_\chi$ and $g_\chi$.

\begin{figure}[t]
    \centering
    \includegraphics[width=0.49\linewidth]{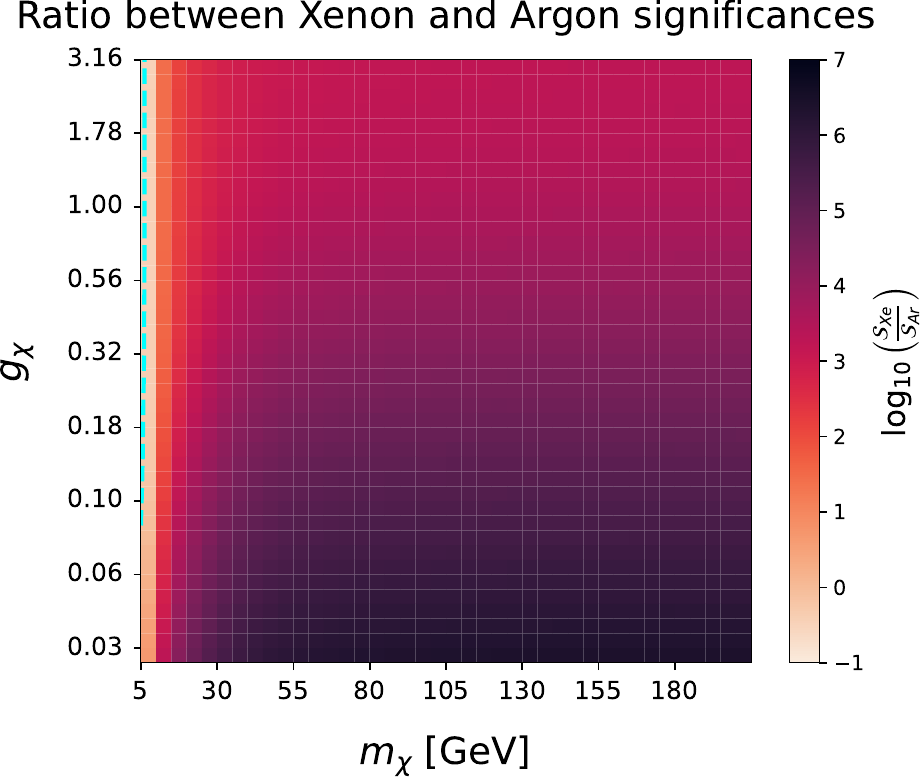}
    \includegraphics[width=0.49\linewidth]{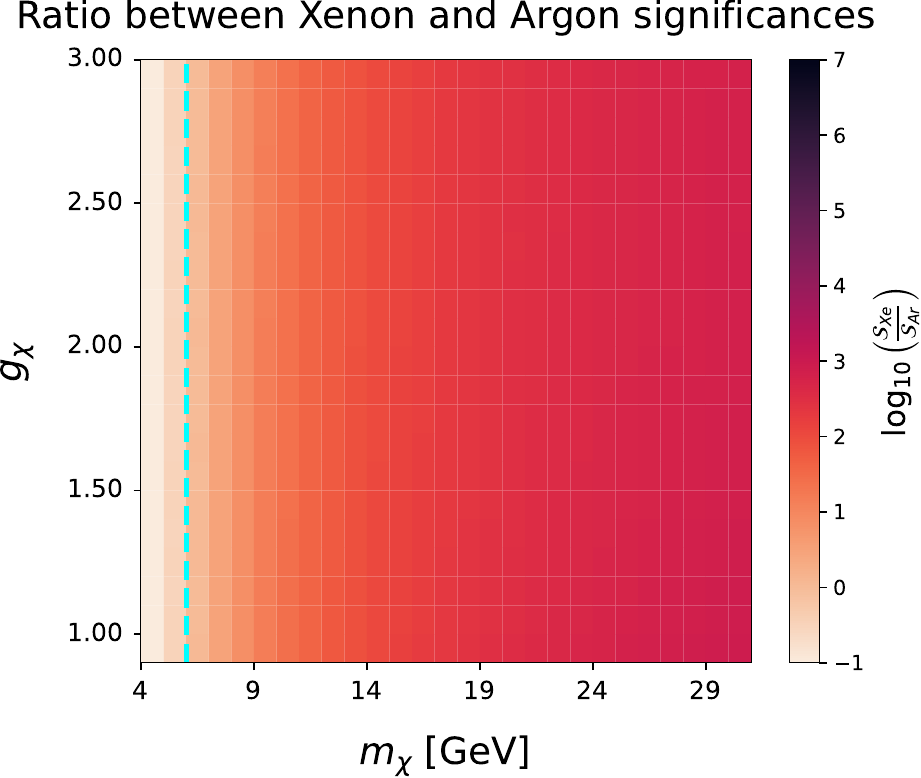}
    \caption{Left: The ratio of the LZ xenon and proposed DS-LM argon signal significances shown in figure~\ref{fig:heatmaps},
    with the same detector configurations as in figure~\ref{fig:heatmaps}.
    Right: Zoom-in of the left panel to the region of lower DM mass and higher DM coupling that favors Ar-based experiments. 
    For most part of the parameter space, the Xe-based experiment has a larger sensitivity than the Ar-based experiment. However, for a small corner of the parameter space, where $m_\chi$ is a few GeV and $g_\chi\sim 1-3$, the Ar-based experiment can reach $\sim 10$ times the  sensitivity of Xe. 
    The near vertical dashed cyan lines on both panels are where the Ar and Xe experiments are of the same sensitivity.
    }
    \label{fig:heatmapRatios}
\end{figure}

To better compare the detection capabilities of the two experiments, in figure~\ref{fig:heatmapRatios} we show the signal significance $\mathcal{S}$ ratio, between LZ and DS-LM.
We note that this plot was created with a 15.3 ton year exposure assumption for LZ and 1 ton year for DS-LM.
The left panel shows this ratio for a large range of $m_\chi$ and $g_\chi$ values, while the right panel zooms in to the region of small $m_\chi$ and large $g_\chi$ values, where the $\mathcal{S}$ ratio is the smallest. 
As $\mathcal{S}\propto\text{(exposure)}^{1/2}$, in general it would take more time or fiducial mass for DS-LM  to reach the same level of sensitivity as the LZ experiment. 
However, in cases where both the DM coupling is large ($g_\chi \sim 1-3$) and the DM is light (a few GeV), Ar-based experiments are favored, as shown in the right panel of figure~\ref{fig:heatmapRatios}. In the most favorable case for Ar, $\mathcal{S}_{\rm Xe}/\mathcal{S}_{\rm Ar} \sim 0.1$, and the DS-LM experiment can reach a sensitivity $\sim 10$ times better than the LZ experiment. 
In this case, we would need significantly more exposure in LZ to achieve a sensitivity similar to that of DS-LM.

\section{Conclusions}
\label{sec:conclusions}

In this work we have compared the capability of Ar and Xe based direct detection experiments for detecting  fermionic DM candidates with interaction mediated by a spin-zero boson with pseudoscalar couplings to the SM fermions in the Lagrangian. At tree-level this type of coupling depends on the spin of the target nucleus and thus would not be detectable in an argon target, since Ar does not have spin. We considered a minimal extension of the simple one mediator model which allows us to compute loop-level contributions to the direct detection event rate that are independent of the nuclear spin.  These contributions are very important at low recoil energies and thus a low energy threshold is essential to detect them.\footnote{After completing our work, we noticed ref.~\cite{Cheek:2023zhv}, which deals with isospin-violation in DM couplings and compares the Xe and Ar sensitivity to SD $(\vec S_\chi \cdot \vec S_N)$  interactions (due to a $\bar f\gamma_\mu \gamma_5 f$ coupling to SM fermions) within two effective SM gauge-invariant models. It finds that Ar becomes sensitive to it at low $E_R$,  due to an $\mathcal{O}_8= (\vec S_\chi \cdot v^{\rm T}) 1_N$ operator ($v^{\rm T}$ is the transverse DM velocity) imposed by gauge invariance.} 

Considering idealized experimental conditions for the two detectors, we showed that with a comparably low energy threshold close to 1 keV, an Ar-based detector could have a similar integrated rate as a Xe-based detector if its exposure is 10 or more times larger than a Xe detector. Considering more realistic experimental conditions based on the DS-LM experiment for Ar and LZ for Xe, we found that in general for large DM masses and small DM couplings we need many times the proposed exposure in the Ar-based DS-LM experiment to reach a sensitivity similar to that of the Xe-based LZ  experiment. However, in a small region of the parameter space, where the DM mass is small (a few GeV) and the DM coupling is large ($g_\chi \sim 1 -3$), the Ar-based experiment is favored and can reach a sensitivity which is $\sim 10$ times better than the Xe-based experiment.

\section*{Acknowledgements}

NB acknowledges the support of the Natural Sciences and Engineering Research Council of Canada (NSERC), funding reference number RGPIN-2020-07138, the NSERC Discovery Launch Supplement, DGECR-2020-00231, and the Canada Research Chairs Program. 
GG was supported in part by the  Department of Energy under Award Number DE-SC0009937. MC was supported in part by the Department of Energy under Award Number DE-
SC000993, and by the World Premier International Research Center Initiative (WPI) MEXT, Japan. ACK and YX were supported in part by the U.S. Department of Energy, Office of Science, Office of High Energy Physics under Award Number DE-SC0025629.



\bibliographystyle{JHEP}
\bibliography{refs}

\providecommand{\href}[2]{#2}\begingroup\raggedright\begin{thebibliography}{10}

\bibitem{XENON:2019rxp}
{\bf XENON} Collaboration, E.~Aprile et~al., {\it {Constraining the
  spin-dependent WIMP-nucleon cross sections with XENON1T}},  {\em Phys. Rev.
  Lett.} {\bf 122} (2019), no.~14 141301,
  [\href{http://arxiv.org/abs/1902.03234}{{\tt arXiv:1902.03234}}].

\bibitem{XENON:2020kmp}
{\bf XENON} Collaboration, E.~Aprile et~al., {\it {Projected WIMP sensitivity
  of the XENONnT dark matter experiment}},  {\em JCAP} {\bf 11} (2020) 031,
  [\href{http://arxiv.org/abs/2007.08796}{{\tt arXiv:2007.08796}}].

\bibitem{LZ:2018qzl}
{\bf LZ} Collaboration, D.~S. Akerib et~al., {\it {Projected WIMP sensitivity
  of the LUX-ZEPLIN dark matter experiment}},  {\em Phys. Rev. D} {\bf 101}
  (2020), no.~5 052002, [\href{http://arxiv.org/abs/1802.06039}{{\tt
  arXiv:1802.06039}}].

\bibitem{LZ:2019sgr}
{\bf LZ} Collaboration, D.~S. Akerib et~al., {\it {The LUX-ZEPLIN (LZ)
  Experiment}},  {\em Nucl. Instrum. Meth. A} {\bf 953} (2020) 163047,
  [\href{http://arxiv.org/abs/1910.09124}{{\tt arXiv:1910.09124}}].

\bibitem{PandaX:2018wtu}
{\bf PandaX} Collaboration, H.~Zhang et~al., {\it {Dark matter direct search
  sensitivity of the PandaX-4T experiment}},  {\em Sci. China Phys. Mech.
  Astron.} {\bf 62} (2019), no.~3 31011,
  [\href{http://arxiv.org/abs/1806.02229}{{\tt arXiv:1806.02229}}].

\bibitem{PandaX-4T:2021bab}
{\bf PandaX-4T} Collaboration, Y.~Meng et~al., {\it {Dark Matter Search Results
  from the PandaX-4T Commissioning Run}},  {\em Phys. Rev. Lett.} {\bf 127}
  (2021), no.~26 261802, [\href{http://arxiv.org/abs/2107.13438}{{\tt
  arXiv:2107.13438}}].

\bibitem{Baudis:2024jnk}
L.~Baudis, {\it {DARWIN/XLZD: A future xenon observatory for dark matter and
  other rare interactions}},  {\em Nucl. Phys. B} {\bf 1003} (2024) 116473,
  [\href{http://arxiv.org/abs/2404.19524}{{\tt arXiv:2404.19524}}].

\bibitem{aalbers2024xlzd}
J.~Aalbers, K.~Abe, M.~Adrover, S.~A. Maouloud, D.~Akerib, A.~Al~Musalhi,
  F.~Alder, L.~Althueser, D.~Amaral, C.~Amarasinghe, et~al., {\it The xlzd
  design book: Towards the next-generation liquid xenon observatory for dark
  matter and neutrino physics},  {\em arXiv preprint arXiv:2410.17137} (2024).

\bibitem{DarkSide-20k:2017zyg}
{\bf DarkSide-20k} Collaboration, C.~E. Aalseth et~al., {\it {DarkSide-20k: A
  20 tonne two-phase LAr TPC for direct dark matter detection at LNGS}},  {\em
  Eur. Phys. J. Plus} {\bf 133} (2018) 131,
  [\href{http://arxiv.org/abs/1707.08145}{{\tt arXiv:1707.08145}}].

\bibitem{ArDM:2017ndf}
{\bf ArDM} Collaboration, J.~Calvo et~al., {\it {Backgrounds and pulse shape
  discrimination in the ArDM liquid argon TPC}},  {\em JCAP} {\bf 12} (2018)
  011, [\href{http://arxiv.org/abs/1712.01932}{{\tt arXiv:1712.01932}}].

\bibitem{DarkSide-50:2023fcw}
{\bf DarkSide-50} Collaboration, P.~Agnes et~al., {\it {Search for low mass
  dark matter in DarkSide-50: the bayesian network approach}},  {\em Eur. Phys.
  J. C} {\bf 83} (2023) 322, [\href{http://arxiv.org/abs/2302.01830}{{\tt
  arXiv:2302.01830}}].

\bibitem{Manthos:2023swh}
{\bf DarkSide-20k} Collaboration, I.~Manthos, {\it {DarkSide-20k: Next
  generation Direct Dark Matter searches with liquid Argon}},  {\em PoS} {\bf
  EPS-HEP2023} (2024) 113, [\href{http://arxiv.org/abs/2312.03597}{{\tt
  arXiv:2312.03597}}].

\bibitem{Aalbers:2022dzr}
J.~Aalbers et~al., {\it {A next-generation liquid xenon observatory for dark
  matter and neutrino physics}},  {\em J. Phys. G} {\bf 50} (2023), no.~1
  013001, [\href{http://arxiv.org/abs/2203.02309}{{\tt arXiv:2203.02309}}].

\bibitem{Bonivento:2024qpn}
W.~M. Bonivento and F.~Terranova, {\it {The science and technology of liquid
  argon detectors}},  \href{http://arxiv.org/abs/2405.01153}{{\tt
  arXiv:2405.01153}}.

\bibitem{DARWIN:2016hyl}
{\bf DARWIN} Collaboration, J.~Aalbers et~al., {\it {DARWIN: towards the
  ultimate dark matter detector}},  {\em JCAP} {\bf 11} (2016) 017,
  [\href{http://arxiv.org/abs/1606.07001}{{\tt arXiv:1606.07001}}].

\bibitem{Galbiati-2018}
C.~Galbiati, ``{Direct dark matter detection with noble gases}.'' Talk at the
  UCLA Dark Matter 2018 Conference.

\bibitem{DarkSide20k:2020ymr}
{\bf DarkSide 20k} Collaboration, P.~Agnes et~al., {\it {Sensitivity of future
  liquid argon dark matter search experiments to core-collapse supernova
  neutrinos}},  {\em JCAP} {\bf 03} (2021) 043,
  [\href{http://arxiv.org/abs/2011.07819}{{\tt arXiv:2011.07819}}].

\bibitem{Gelmini:2018ogy}
G.~B. Gelmini, V.~Takhistov, and S.~J. Witte, {\it {Casting a Wide Signal Net
  with Future Direct Dark Matter Detection Experiments}},  {\em JCAP} {\bf 07}
  (2018) 009, [\href{http://arxiv.org/abs/1804.01638}{{\tt arXiv:1804.01638}}].
  [Erratum: JCAP 02, E02 (2019)].

\bibitem{Arcadi:2017wqi}
G.~Arcadi, M.~Lindner, F.~S. Queiroz, W.~Rodejohann, and S.~Vogl, {\it
  {Pseudoscalar Mediators: A WIMP model at the Neutrino Floor}},  {\em JCAP}
  {\bf 03} (2018) 042, [\href{http://arxiv.org/abs/1711.02110}{{\tt
  arXiv:1711.02110}}].

\bibitem{Bell:2018zra}
N.~F. Bell, G.~Busoni, and I.~W. Sanderson, {\it {Loop Effects in Direct
  Detection}},  {\em JCAP} {\bf 08} (2018) 017,
  [\href{http://arxiv.org/abs/1803.01574}{{\tt arXiv:1803.01574}}]. [Erratum:
  JCAP 01, E01 (2019)].

\bibitem{Li:2018qip}
T.~Li, {\it {Revisiting the direct detection of dark matter in simplified
  models}},  {\em Phys. Lett. B} {\bf 782} (2018) 497--502,
  [\href{http://arxiv.org/abs/1804.02120}{{\tt arXiv:1804.02120}}].

\bibitem{Abe:2018emu}
T.~Abe, M.~Fujiwara, and J.~Hisano, {\it {Loop corrections to dark matter
  direct detection in a pseudoscalar mediator dark matter model}},  {\em JHEP}
  {\bf 02} (2019) 028, [\href{http://arxiv.org/abs/1810.01039}{{\tt
  arXiv:1810.01039}}].

\bibitem{Ertas:2019dew}
F.~Ertas and F.~Kahlhoefer, {\it {Loop-induced direct detection signatures from
  CP-violating scalar mediators}},  {\em JHEP} {\bf 06} (2019) 052,
  [\href{http://arxiv.org/abs/1902.11070}{{\tt arXiv:1902.11070}}].

\bibitem{Freytsis:2010ne}
M.~Freytsis and Z.~Ligeti, {\it {On dark matter models with uniquely
  spin-dependent detection possibilities}},  {\em Phys. Rev. D} {\bf 83} (2011)
  115009, [\href{http://arxiv.org/abs/1012.5317}{{\tt arXiv:1012.5317}}].

\bibitem{Dienes:2013xya}
K.~R. Dienes, J.~Kumar, B.~Thomas, and D.~Yaylali, {\it {Overcoming Velocity
  Suppression in Dark-Matter Direct-Detection Experiments}},  {\em Phys. Rev.
  D} {\bf 90} (2014), no.~1 015012, [\href{http://arxiv.org/abs/1312.7772}{{\tt
  arXiv:1312.7772}}].

\bibitem{Boehm:2014hva}
C.~Boehm, M.~J. Dolan, C.~McCabe, M.~Spannowsky, and C.~J. Wallace, {\it
  {Extended gamma-ray emission from Coy Dark Matter}},  {\em JCAP} {\bf 05}
  (2014) 009, [\href{http://arxiv.org/abs/1401.6458}{{\tt arXiv:1401.6458}}].

\bibitem{Ipek:2014gua}
S.~Ipek, D.~McKeen, and A.~E. Nelson, {\it {A Renormalizable Model for the
  Galactic Center Gamma Ray Excess from Dark Matter Annihilation}},  {\em Phys.
  Rev. D} {\bf 90} (2014), no.~5 055021,
  [\href{http://arxiv.org/abs/1404.3716}{{\tt arXiv:1404.3716}}].

\bibitem{No:2015xqa}
J.~M. No, {\it {Looking through the pseudoscalar portal into dark matter: Novel
  mono-Higgs and mono-Z signatures at the LHC}},  {\em Phys. Rev. D} {\bf 93}
  (2016), no.~3 031701, [\href{http://arxiv.org/abs/1509.01110}{{\tt
  arXiv:1509.01110}}].

\bibitem{Goncalves:2016iyg}
D.~Goncalves, P.~A.~N. Machado, and J.~M. No, {\it {Simplified Models for Dark
  Matter Face their Consistent Completions}},  {\em Phys. Rev. D} {\bf 95}
  (2017), no.~5 055027, [\href{http://arxiv.org/abs/1611.04593}{{\tt
  arXiv:1611.04593}}].

\bibitem{Bauer:2017ota}
M.~Bauer, U.~Haisch, and F.~Kahlhoefer, {\it {Simplified dark matter models
  with two Higgs doublets: I. Pseudoscalar mediators}},  {\em JHEP} {\bf 05}
  (2017) 138, [\href{http://arxiv.org/abs/1701.07427}{{\tt arXiv:1701.07427}}].

\bibitem{Bauer:2017fsw}
M.~Bauer, M.~Klassen, and V.~Tenorth, {\it {Universal properties of
  pseudoscalar mediators in dark matter extensions of 2HDMs}},  {\em JHEP} {\bf
  07} (2018) 107, [\href{http://arxiv.org/abs/1712.06597}{{\tt
  arXiv:1712.06597}}].

\bibitem{LHCDarkMatterWorkingGroup:2018ufk}
{\bf LHC Dark Matter Working Group} Collaboration, T.~Abe et~al., {\it {LHC
  Dark Matter Working Group: Next-generation spin-0 dark matter models}},  {\em
  Phys. Dark Univ.} {\bf 27} (2020) 100351,
  [\href{http://arxiv.org/abs/1810.09420}{{\tt arXiv:1810.09420}}].

\bibitem{DAmbrosio:2002vsn}
G.~D'Ambrosio, G.~F. Giudice, G.~Isidori, and A.~Strumia, {\it {Minimal flavor
  violation: An Effective field theory approach}},  {\em Nucl. Phys. B} {\bf
  645} (2002) 155--187, [\href{http://arxiv.org/abs/hep-ph/0207036}{{\tt
  hep-ph/0207036}}].

\bibitem{Bishara:2018vix}
F.~Bishara, J.~Brod, B.~Grinstein, and J.~Zupan, {\it {Renormalization Group
  Effects in Dark Matter Interactions}},  {\em JHEP} {\bf 03} (2020) 089,
  [\href{http://arxiv.org/abs/1809.03506}{{\tt arXiv:1809.03506}}].

\bibitem{Passarino:1978jh}
G.~Passarino and M.~J.~G. Veltman, {\it {One Loop Corrections for e+ e-
  Annihilation Into mu+ mu- in the Weinberg Model}},  {\em Nucl. Phys. B} {\bf
  160} (1979) 151--207.

\bibitem{Abe:2015rja}
T.~Abe and R.~Sato, {\it {Quantum corrections to the spin-independent cross
  section of the inert doublet dark matter}},  {\em JHEP} {\bf 03} (2015) 109,
  [\href{http://arxiv.org/abs/1501.04161}{{\tt arXiv:1501.04161}}].

\bibitem{Patel:2015tea}
H.~H. Patel, {\it {Package-X: A Mathematica package for the analytic
  calculation of one-loop integrals}},  {\em Comput. Phys. Commun.} {\bf 197}
  (2015) 276--290, [\href{http://arxiv.org/abs/1503.01469}{{\tt
  arXiv:1503.01469}}].

\bibitem{Patel:2016fam}
H.~H. Patel, {\it {Package-X 2.0: A Mathematica package for the analytic
  calculation of one-loop integrals}},  {\em Comput. Phys. Commun.} {\bf 218}
  (2017) 66--70, [\href{http://arxiv.org/abs/1612.00009}{{\tt
  arXiv:1612.00009}}].

\bibitem{Denner:2016kdg}
A.~Denner, S.~Dittmaier, and L.~Hofer, {\it {Collier: a fortran-based Complex
  One-Loop LIbrary in Extended Regularizations}},  {\em Comput. Phys. Commun.}
  {\bf 212} (2017) 220--238, [\href{http://arxiv.org/abs/1604.06792}{{\tt
  arXiv:1604.06792}}].

\bibitem{Bishara:2017pfq}
F.~Bishara, J.~Brod, B.~Grinstein, and J.~Zupan, {\it {From quarks to nucleons
  in dark matter direct detection}},  {\em JHEP} {\bf 11} (2017) 059,
  [\href{http://arxiv.org/abs/1707.06998}{{\tt arXiv:1707.06998}}].

\bibitem{Haxton:2024lyc}
W.~Haxton, K.~McElvain, T.~Menzo, E.~Rule, and J.~Zupan, {\it {Effective theory
  tower for $\mu\rightarrow e$ conversion}},
  \href{http://arxiv.org/abs/2406.13818}{{\tt arXiv:2406.13818}}.

\bibitem{Fan:2010gt}
J.~Fan, M.~Reece, and L.-T. Wang, {\it {Non-relativistic effective theory of
  dark matter direct detection}},  {\em JCAP} {\bf 11} (2010) 042,
  [\href{http://arxiv.org/abs/1008.1591}{{\tt arXiv:1008.1591}}].

\bibitem{Fitzpatrick:2012ix}
A.~L. Fitzpatrick, W.~Haxton, E.~Katz, N.~Lubbers, and Y.~Xu, {\it {The
  Effective Field Theory of Dark Matter Direct Detection}},  {\em JCAP} {\bf
  02} (2013) 004, [\href{http://arxiv.org/abs/1203.3542}{{\tt
  arXiv:1203.3542}}].

\bibitem{Fitzpatrick:2012ib}
A.~L. Fitzpatrick, W.~Haxton, E.~Katz, N.~Lubbers, and Y.~Xu, {\it {Model
  Independent Direct Detection Analyses}},
  \href{http://arxiv.org/abs/1211.2818}{{\tt arXiv:1211.2818}}.

\bibitem{Anand:2013yka}
N.~Anand, A.~L. Fitzpatrick, and W.~C. Haxton, {\it {Weakly interacting massive
  particle-nucleus elastic scattering response}},  {\em Phys. Rev. C} {\bf 89}
  (2014), no.~6 065501, [\href{http://arxiv.org/abs/1308.6288}{{\tt
  arXiv:1308.6288}}].

\bibitem{Dent:2015zpa}
J.~B. Dent, L.~M. Krauss, J.~L. Newstead, and S.~Sabharwal, {\it {General
  analysis of direct dark matter detection: From microphysics to observational
  signatures}},  {\em Phys. Rev. D} {\bf 92} (2015), no.~6 063515,
  [\href{http://arxiv.org/abs/1505.03117}{{\tt arXiv:1505.03117}}].

\bibitem{Drukier:1986tm}
A.~K. Drukier, K.~Freese, and D.~N. Spergel, {\it {Detecting Cold Dark Matter
  Candidates}},  {\em Phys. Rev. D} {\bf 33} (1986) 3495--3508.

\bibitem{Bozorgnia:2016ogo}
N.~Bozorgnia, F.~Calore, M.~Schaller, M.~Lovell, G.~Bertone, C.~S. Frenk, R.~A.
  Crain, J.~F. Navarro, J.~Schaye, and T.~Theuns, {\it Simulated milky way
  analogues: implications for dark matter direct searches},  {\em JCAP} {\bf
  05} (2016) 024, [\href{http://arxiv.org/abs/1601.04707}{{\tt
  arXiv:1601.04707}}].

\bibitem{Kelso:2016qqj}
C.~Kelso, C.~Savage, M.~Valluri, K.~Freese, G.~S. Stinson, and J.~Bailin, {\it
  The impact of baryons on the direct detection of dark matter},  {\em JCAP}
  {\bf 08} (2016) 071, [\href{http://arxiv.org/abs/1601.04725}{{\tt
  arXiv:1601.04725}}].

\bibitem{Sloane:2016kyi}
J.~D. Sloane, M.~R. Buckley, A.~M. Brooks, and F.~Governato, {\it Assessing
  astrophysical uncertainties in direct detection with galaxy simulations},
  {\em Astrophys.J.} {\bf 831} (2016) 93,
  [\href{http://arxiv.org/abs/1601.05402}{{\tt arXiv:1601.05402}}].

\bibitem{Poole-McKenzie:2020dbo}
R.~Poole-McKenzie, A.~S. Font, B.~Boxer, I.~G. McCarthy, S.~Burdin, S.~G.
  Stafford, and S.~T. Brown, {\it {Informing dark matter direct detection
  limits with the ARTEMIS simulations}},  {\em JCAP} {\bf 11} (2020) 016,
  [\href{http://arxiv.org/abs/2006.15159}{{\tt arXiv:2006.15159}}].

\bibitem{Reynoso-Cordova:2024xqz}
J.~Reynoso-Cordova, N.~Bozorgnia, and M.-C. Piro, {\it {The Large Magellanic
  Cloud: expanding the low-mass parameter space of dark matter direct
  detection}},  {\em JCAP} {\bf 12} (2024) 037,
  [\href{http://arxiv.org/abs/2409.09119}{{\tt arXiv:2409.09119}}].

\bibitem{Smith-Orlik:2023kyl}
A.~Smith-Orlik et~al., {\it {The impact of the Large Magellanic Cloud on dark
  matter direct detection signals}},  {\em JCAP} {\bf 10} (2023) 070,
  [\href{http://arxiv.org/abs/2302.04281}{{\tt arXiv:2302.04281}}].

\bibitem{Besla:2019xbx}
G.~Besla, A.~Peter, and N.~Garavito-Camargo, {\it {The highest-speed local dark
  matter particles come from the Large Magellanic Cloud}},  {\em JCAP} {\bf 11}
  (2019) 013, [\href{http://arxiv.org/abs/1909.04140}{{\tt arXiv:1909.04140}}].

\bibitem{Donaldson:2021byu}
K.~Donaldson, M.~S. Petersen, and J.~Pe\~narrubia, {\it {Effects on the local
  dark matter distribution due to the large magellanic cloud}},  {\em Mon. Not.
  Roy. Astron. Soc.} {\bf 513} (2022), no.~1 46--51,
  [\href{http://arxiv.org/abs/2111.15440}{{\tt arXiv:2111.15440}}].

\bibitem{Reid:2009nj}
M.~J. Reid et~al., {\it {Trigonometric Parallaxes of Massive Star Forming
  Regions: VI. Galactic Structure, Fundamental Parameters and Non-Circular
  Motions}},  {\em Astrophys. J.} {\bf 700} (2009) 137--148,
  [\href{http://arxiv.org/abs/0902.3913}{{\tt arXiv:0902.3913}}].

\bibitem{Bovy:2009dr}
J.~Bovy, D.~W. Hogg, and H.-W. Rix, {\it {Galactic masers and the Milky Way
  circular velocity}},  {\em Astrophys. J.} {\bf 704} (2009) 1704--1709,
  [\href{http://arxiv.org/abs/0907.5423}{{\tt arXiv:0907.5423}}].

\bibitem{Smith:2006ym}
M.~C. Smith et~al., {\it {The RAVE Survey: Constraining the Local Galactic
  Escape Speed}},  {\em Mon. Not. Roy. Astron. Soc.} {\bf 379} (2007) 755--772,
  [\href{http://arxiv.org/abs/astro-ph/0611671}{{\tt astro-ph/0611671}}].

\bibitem{Schoenrich:2009bx}
R.~Schoenrich, J.~Binney, and W.~Dehnen, {\it {Local Kinematics and the Local
  Standard of Rest}},  {\em Mon. Not. Roy. Astron. Soc.} {\bf 403} (2010) 1829,
  [\href{http://arxiv.org/abs/0912.3693}{{\tt arXiv:0912.3693}}].

\bibitem{DarkSide:2018bpj}
{\bf DarkSide} Collaboration, P.~Agnes et~al., {\it {Low-Mass Dark Matter
  Search with the DarkSide-50 Experiment}},  {\em Phys. Rev. Lett.} {\bf 121}
  (2018), no.~8 081307, [\href{http://arxiv.org/abs/1802.06994}{{\tt
  arXiv:1802.06994}}].

\bibitem{Wada:2024xqq}
{\bf Global Argon Dark Matter} Collaboration, M.~Wada, {\it {DarkSide-LowMass:
  Sensitivity Projections for a New Detector Designed for Light Dark Matter
  Searches}},  {\em PoS} {\bf TAUP2023} (2024) 057.

\bibitem{GlobalArgonDarkMatter:2022ppc}
{\bf Global Argon Dark Matter} Collaboration, P.~Agnes et~al., {\it
  {Sensitivity projections for a dual-phase argon TPC optimized for light dark
  matter searches through the ionization channel}},  {\em Phys. Rev. D} {\bf
  107} (2023), no.~11 112006, [\href{http://arxiv.org/abs/2209.01177}{{\tt
  arXiv:2209.01177}}].

\bibitem{Fan:2016ymy}
A.~Fan, {\em {Results from the DarkSide-50 Dark Matter Experiment}}.
\newblock PhD thesis, UCLA, 2016.

\bibitem{Cadeddu:2018osv}
M.~Cadeddu, {\em {DarkSide-20k sensitivity, directional dark matter detection
  and the role of coherent elastic neutrino-nucleus scattering background}}.
\newblock PhD thesis, Cagliari U., 2018.

\bibitem{DarkSide:2021bnz}
{\bf DarkSide} Collaboration, P.~Agnes et~al., {\it {Calibration of the liquid
  argon ionization response to low energy electronic and nuclear recoils with
  DarkSide-50}},  {\em Phys. Rev. D} {\bf 104} (2021), no.~8 082005,
  [\href{http://arxiv.org/abs/2107.08087}{{\tt arXiv:2107.08087}}].

\bibitem{Mount:2017qzi}
B.~J. Mount et~al., {\it {LUX-ZEPLIN (LZ) Technical Design Report}},
  \href{http://arxiv.org/abs/1703.09144}{{\tt arXiv:1703.09144}}.

\bibitem{LZ:2022lsv}
{\bf LZ} Collaboration, J.~Aalbers et~al., {\it {First Dark Matter Search
  Results from the LUX-ZEPLIN (LZ) Experiment}},  {\em Phys. Rev. Lett.} {\bf
  131} (2023), no.~4 041002, [\href{http://arxiv.org/abs/2207.03764}{{\tt
  arXiv:2207.03764}}].

\bibitem{Szydagis:2011tk}
M.~Szydagis, N.~Barry, K.~Kazkaz, J.~Mock, D.~Stolp, M.~Sweany, M.~Tripathi,
  S.~Uvarov, N.~Walsh, and M.~Woods, {\it {NEST: A Comprehensive Model for
  Scintillation Yield in Liquid Xenon}},  {\em JINST} {\bf 6} (2011) P10002,
  [\href{http://arxiv.org/abs/1106.1613}{{\tt arXiv:1106.1613}}].

\bibitem{Cheek:2023zhv}
A.~Cheek, D.~D. Price, and E.~M. Sandford, {\it {Isospin-violating dark matter
  at liquid noble detectors: new constraints, future projections, and an
  exploration of target complementarity}},  {\em Eur. Phys. J. C} {\bf 83}
  (2023), no.~10 914, [\href{http://arxiv.org/abs/2302.05458}{{\tt
  arXiv:2302.05458}}].

\end{thebibliography}\endgroup
\end{document}